\documentclass[aps,prl,preprint,showpacs,tightenlines]{revtex4-1}

\usepackage{graphicx}

\usepackage{color}

\begin{document}


\title{Topological Constraints on Magnetic Relaxation} 



\author{A. R. Yeates}
 \email{anthony@maths.dundee.ac.uk}
\author{G. Hornig}
\author{A. L. Wilmot-Smith}
\affiliation{Division of Mathematics, University of Dundee, Dundee, DD1 4HN, UK}


\date{\today}

\begin{abstract}
The final state of turbulent magnetic relaxation in a reversed field pinch is well explained by Taylor's hypothesis. However, recent resistive-magnetohydrodynamic simulations of the relaxation of braided solar coronal loops have led to relaxed fields far from the Taylor state, despite the conservation of helicity. We point out the existence of an additional topological invariant in any flux tube with non-zero field: the topological degree of the field line mapping. We conjecture that this constrains the relaxation, explaining why only one of three example simulations reaches the Taylor state.
\end{abstract}


\maketitle 


The landmark paper of J. B. Taylor \cite{taylor1974,*taylor2000} showed how the final, relaxed, magnetic field in a turbulent plasma experiment could be predicted theoretically by assuming conservation of the total magnetic helicity $H=\int_V\mathbf{A}\cdot\mathbf{B}\,d^3x,$ where $\mathbf{B}=\nabla\times\mathbf{A}$. Taylor hypothesized that, in a turbulent plasma with small but non-vanishing resistivity, $H$ is the only relevant constraint on the relaxation. The resulting minimum energy state is then a linear force-free field \cite{woltjer1958}, $\nabla\times\mathbf{B}=\alpha\mathbf{B}$, where $\alpha$ is a constant depending on the value of $H$.

The success of this theory in predicting the final state of relaxation in a reversed field pinch has led to speculation that Taylor's hypothesis might apply more generally, for example, to magnetic structures in the Sun's atmosphere \cite{heyvaerts1984,dixon1989,vekstein1993,browning2008}. Here, the nature of the final state is of great interest because it limits the magnetic energy available for conversion to heat during the relaxation. This is key to understanding how magnetic fields produce the extreme coronal temperatures in the Sun and other stars. To apply Taylor's hypothesis in such magnetically open domains, $H$ is replaced by the relative helicity $H_R$ with respect to a reference field \cite{berger1984,dixon1989}, typically a potential field. A local domain, such as an isolated loop, is taken (the corona cannot be globally a linear force-free field) and the hypothesis has met with some notable success \cite{browning2008}. However, counter-examples are known \cite{amari2000,pontin2010} and it now seems likely that constraints beyond the conservation of global magnetic helicity  may be important for general relaxation events.  For example, while $H_R$ measures only the second-order linkage of field lines, higher-order invariants may play a role \cite{bhattacharjee1980,*bhattacharjee1982,ruzmaikin1994,hornig2002,low2006} (although this remains under debate \cite{delsordo2010}).

\begin{figure}
\includegraphics[width=0.5\textwidth]{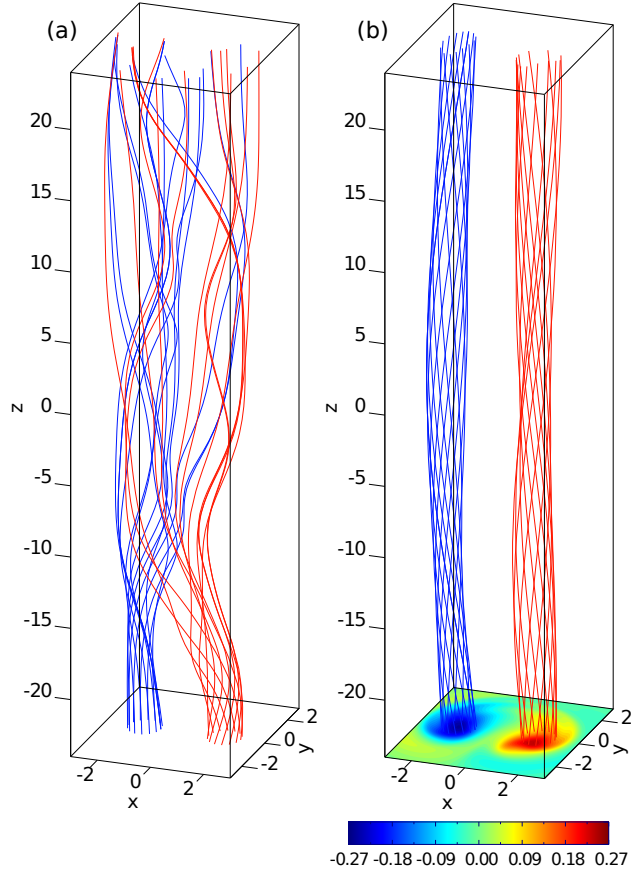}%
\caption{\label{fig:sd256} (Color online) Magnetic field lines in the original braiding experiment \cite{wilmotsmith2010} for (a) the initial state at $t=0$, and (b) the relaxed state at $t=290$ (in units of the Alfv\'{e}n time). Field lines are traced from the same starting points in each case. In (b), color contours show $\alpha=\mathbf{j}\cdot\mathbf{B}/B^2$ on the $S_0$ plane.}%
\end{figure}

This Letter presents a conjecture on one possible topological constraint and was motivated by
recent resistive-magnetohydrodynamic simulations of the relaxation of a magnetic flux tube, intended to model a solar coronal loop. In all of our experiments the initial flux tube contains a braided magnetic field of non-trivial topology but with no net current or helicity \cite{wilmotsmith2010,pontin2010}. Despite undergoing a turbulent relaxation with myriad small-scale current structures and magnetic reconnections, on a timescale short compared to the resistive timescale, the magnetic field relaxes to a final state which directly contradicts that predicted by Taylor's hypothesis. Although force-free, the final state is not {\it linear} force-free, because the coefficient $\alpha$ varies strongly across field lines. For the first experiment, where we started from a field modeled on the pigtail braid, the current is concentrated into two flux tubes of opposite twist (sign of the parallel current). The initial and final states of this braiding experiment are shown in Fig.~\ref{fig:sd256}. If $H_{R} $ were the only relevant constraint on relaxation then the final state would be a near-uniform, vertical, potential field ($\mathbf{B}=\nabla\psi$), because the initial configuration is chosen such that $H_R=0$. Note that $H_R$ is still well conserved during the turbulent relaxation, so the departure of the final state from the Taylor prediction is not due to changes in $H_R$. Rather, the result seems to indicate the presence of additional constraint(s).

In this Letter, we propose that such a constraint is given by a property of the field line mapping called its {\it topological degree}. This is a topological property that can be defined for any continuous mapping from one compact manifold to another. In the braiding experiment, the mapping of magnetic field lines from the lower boundary $S_0$ (where they enter the domain) to the upper boundary $S_1$ (where they exit) forms such a mapping. This is because (a) all field lines connect, in the same direction, these two boundaries (the vertical flux through any horizontal cross-section is constant through the domain), and (b) the magnetic field is everywhere non-zero ($\mathbf{B}\neq 0$) within the domain. Although the fields considered in this paper are defined in rectangular, Cartesian domains such that the flux tube is aligned along the $z$-axis and the boundaries are rectangles in the $(x,y)$ plane, the ideas are more generally applicable because every flux tube of any shape with uni-directional field $\mathbf{B}\neq 0$ is topologically equivalent to a straight cylinder \cite{seehafer1986}. 

To define the topological degree of the field line mapping, let $\mathbf{F}(\mathbf{x_0}) \in S_1$ be the end-point of the field line starting at the point $\mathbf{x}_0 \in S_0$. The field line mapping will have one or more {\it periodic orbits} $\mathbf{x}_0^p \in S_0$ where $\mathbf{F}(\mathbf{x}_0^p) = \mathbf{x}_0^p$ \footnote{For a general flux tube this identification is given by the field line mapping of the reference potential field satisfying the same boundary conditions on $S_0$ and $S_1$ as $\mathbf{B}$. This is the same reference field used for calculating $H_R$.}. As a fixed point of the mapping $\mathbf{F}-\mathbf{I}$ (where $\mathbf{I}$ denotes the identity map), each $\mathbf{x}_0^p$ is characterized by a {\it fixed point index}, defined as the local Brouwer degree of the mapping (see e.g.~\cite{frankel2004}). This index takes integer values, $\pm 1$ for generic, structurally stable, isolated periodic orbits. The case $+1$ corresponds to an elliptic null point of the local linearisation of $\mathbf{F}-\mathbf{I}$, or an elliptic periodic orbit, while $-1$ indicates a hyperbolic periodic orbit. The sum
\begin{equation}
T=\sum_{\mathbf{x}_0^p}\textrm{index}(\mathbf{x}_0^p)
\label{eqn:td}
\end{equation}
over all isolated periodic orbits is called the Lefschetz number or {\it topological degree} of the mapping $\mathbf{F}$. By the Lefschetz-Hopf theorem \cite{brown1971} it is a conserved quantity, providing that no periodic orbits cross the side boundary of the domain. Periodic orbits can therefore be created or annihilated only in pairs of opposite index.

\begin{figure}
\includegraphics[width=0.5\textwidth]{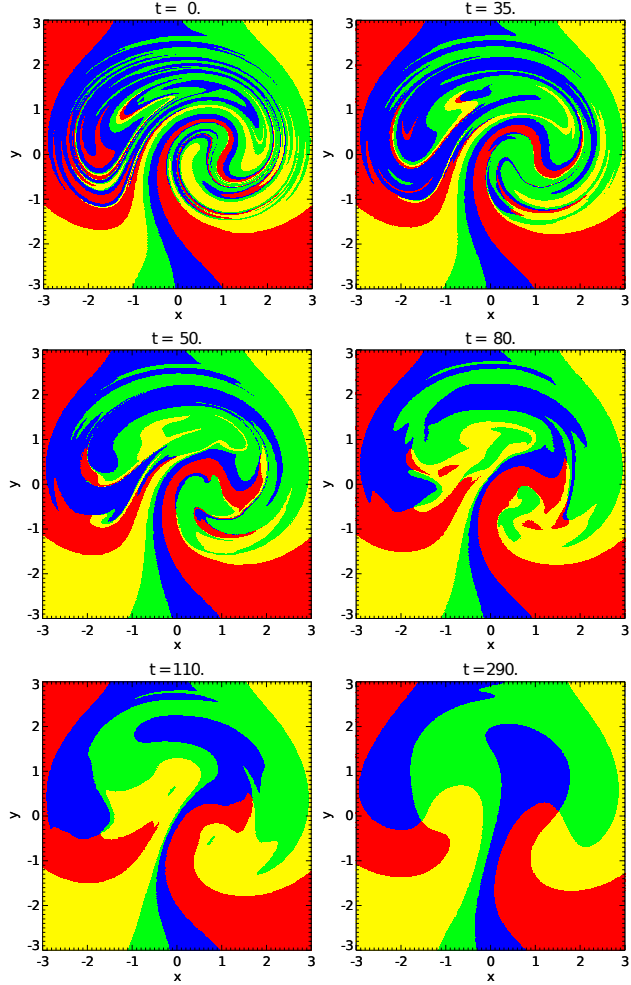}%
\caption{\label{fig:sd256maps} (Color online) Sequence of color maps for the original braiding experiment \cite{wilmotsmith2010}, where total degree $T=2$.} 
\end{figure}

The topological degree $T$ may be computed by evaluating the Kronecker integral around the boundary of $S_0$ \cite{polymilis2003}, or by other numerical methods \cite{stenger1978,*kearfott1979}. Here, we adopt the graphical {\it color map} technique of \cite{polymilis2003}. For example, Fig.~\ref{fig:sd256maps} shows the colour maps at various times in the braiding simulation.
Every point $\mathbf{x}_0=(x_0,y_0)$ on the lower boundary of the domain ($S_0$) is assigned one of four colours, according to its field line mapping $\mathbf{F}(\mathbf{x}_0)=(F_x,F_y)$ to the upper boundary $S_1$. We use {\it red} if $F_x>x_0$ and $F_y>y_0$; {\it yellow} if $F_x<x_0$ and $F_y>y_0$; {\it green} if $F_x<x_0$ and $F_y<y_0$; and {\it blue} if $F_x>x_0$ and $F_y<y_0$.
On the resulting color map, periodic orbits correspond either to red-green or yellow-blue boundaries; isolated, generic periodic orbits are points where all four colours meet. The index of an isolated periodic orbit may be read from the sequence of colors passed through in an anti-clockwise direction around a small circle around the point. An elliptic orbit (index $1$) has red-yellow-green-blue (r-y-g-b), while a hyperbolic orbit (index $-1$) has r-b-g-y. To determine $T$, either sum the individual indices or simply record the sequence of colors around the boundary of $S_0$. For example, in Fig.~\ref{fig:sd256maps}(a),
the initial state for the braiding experiment, we find 12 periodic orbits with index 1, and 10 with index -1, giving a net topological degree of 2. This corresponds with the (anti-clockwise) sequence of colors on the boundary of r-y-g-b-r-y-g-b .

The later color maps in Fig.~\ref{fig:sd256maps} for the braiding experiment show that, during the turbulent relaxation, changes in the magnetic topology by reconnection lead to the annihilation of periodic orbits in pairs of index $+1$ and $-1$. This is consistent with (\ref{eqn:td}) as the reconnection processes occur strictly in the interior of the domain such that $T$ remains unchanged. Eventually  only two elliptic periodic orbits remain in the final relaxed state, lying within the two flux tubes of opposite twist shown in Fig.~\ref{fig:sd256}(b). This leads us to suggest an explanation for the discrepancy between this relaxed state and that predicted by Taylor theory: for a turbulent relaxation in a magnetic flux tube that leaves the side boundary fixed, {\it the topological degree $T$ of the initial field line mapping is an additional topological constraint that can prevent the system from reaching the Taylor state}.

\begin{figure}
\includegraphics[width=0.35\textwidth]{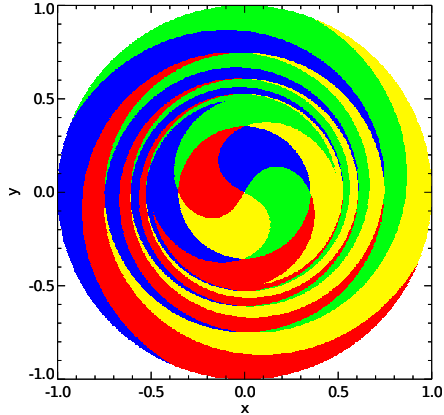}%
\caption{\label{fig:hood} (Color online) Color map for the initial magnetic field of the Browning et al. \cite{browning2008} simulation (case 1).}%
\end{figure}

We now consider another two resistive-MHD simulations of turbulent relaxation in magnetic flux tubes which further support our degree conjecture. The first is a simulation by Browning et al. \cite{browning2008} where the Taylor hypothesis succeeds in predicting the relaxed state. Using their expressions for $\mathbf{B}$, we have computed the color map (shown in Fig.~\ref{fig:hood}) of the initial force-free magnetic field in their case 1. The field consists of an axisymmetric twisted magnetic flux tube of unit radius, with a piecewise-constant profile of $\alpha$ with radius. Fig.~\ref{fig:hood} shows that this initial configuration has a single isolated periodic orbit at the origin. In addition, there are four rational surfaces on which field lines have a winding number that exactly divides $2\pi$, forming four continuous circles of periodic orbits in the $S_0$ plane. However, it is easy to show that such rational surfaces do not contribute to the total degree, so $T=1$, in accordance with the pattern of colors around the tube boundary, $r=1$.

The initial flux tube in the Browning et al. simulation is kink unstable, and a turbulent relaxation is triggered by an initial velocity perturbation. The system relaxes to a force-free equilibrium on a comparable timescale to our original braiding experiment. However, the difference here is that this final equilibrium is a constant-$\alpha$ ``Lundquist'' magnetic field, in accordance with expectations from Taylor theory \footnote{There are inevitably numerical differences owing to computational limitations, and to the fact that helicity is not perfectly conserved: for example, the actual best-fit value of $\alpha$ in the final state differs from that predicted \cite{browning2008}}. The simulation still supports our degree conjecture because the final state retains a total degree $T=1$, with a single twisted flux tube. The difference between this and our braiding experiment is that here the Taylor state is compatible with the degree of the initial field.

\begin{figure}
\includegraphics[width=0.5\textwidth]{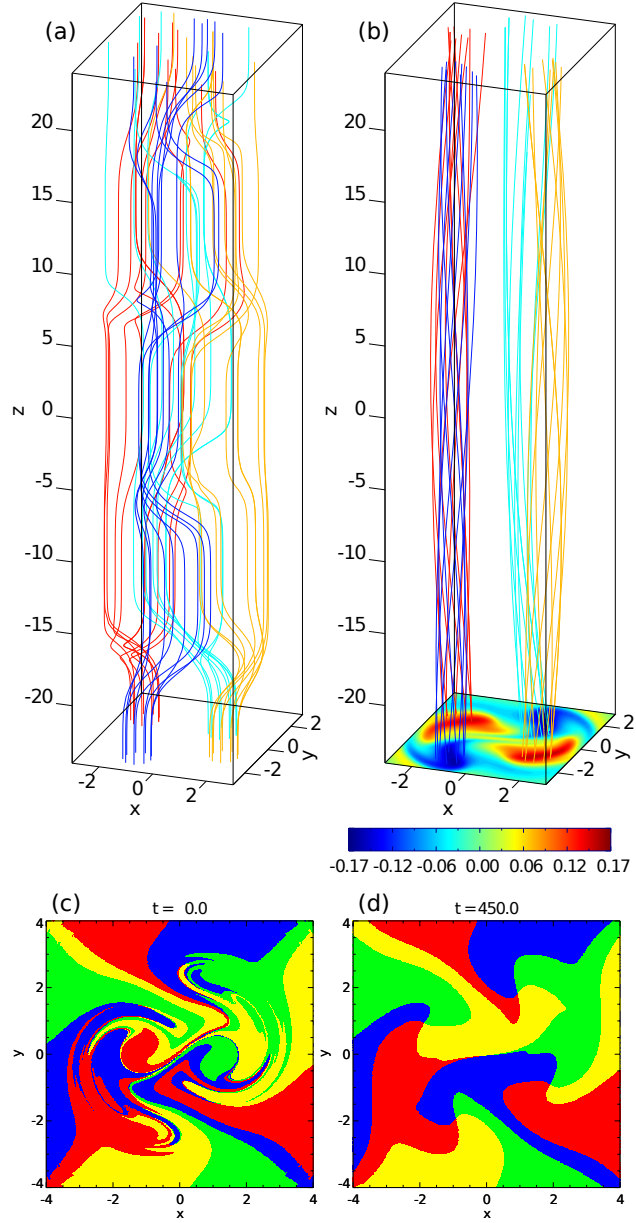}%
\caption{\label{fig:silver} (Color online) The $T=3$ experiment, showing the initial (a) and final (b) magnetic fields (with field lines traced from the same starting points in each case), and the initial (c) and final (d) color maps. In (b), color contours show $\alpha$ on the $S_0$ plane.}%
\end{figure}

Our final example is a new simulation of an initially braided magnetic field with degree $T=3$, solving the same resistive-MHD equations as in our earlier braiding simulation \cite{wilmotsmith2010}. The original initial condition with $T=2$ was modeled on the pigtail braid (see Fig.~2 of Ref.~\cite{wilmotsmith2009}), and was formed from a uniform vertical field super-imposed with six localized ``twist'' regions (Eq.~2 of \cite{wilmotsmith2009}) arranged in two columns. The $T=2$ property arose from the arrangement of the positive twist regions in one column and the negative regions in the other. We obtain a topologically stable $T=3$ configuration by using four columns of twist regions arranged on a circle, where each contribute an elliptic periodic orbit. The total degree is 3 because the remainder of the field contributes a net degree of $-1$. The initial field we used is shown in Fig.~\ref{fig:silver}(a). In the notation of Ref.~\cite{wilmotsmith2009}, the twist regions have parameters $\mathbf{c}_1=(r_0,0,-18,1,\sqrt{2},2)$, $\mathbf{c}_2=(-r_0,0,-18,1,\sqrt{2},2)$, $\mathbf{c}_3=(0,r_0,-6,-1,\sqrt{2},2)$, $\mathbf{c}_4=(0,-r_0,-6,-1,\sqrt{2},2)$,  $\mathbf{c}_5=(r_0,0,6,1,\sqrt{2},2)$, $\mathbf{c}_6=(-r_0,0,6,1,\sqrt{2},2)$ , $\mathbf{c}_7=(0,r_0,18,-1,\sqrt{2},2)$, and  $\mathbf{c}_8=(0,-r_0,18,-1,\sqrt{2},2)$, where $r_0=1.27$. Note that the total helicity of this field vanishes, as in the $T=2$ experiment. The color map for this initial field (Fig.~\ref{fig:silver}c) confirms that $T=3$.

The $T=3$ initial condition leads again to a turbulent relaxation, reaching a relaxed state on a similar timescale to the $T=2$ experiment. The final color map is shown in Fig.~\ref{fig:silver}(d). As in the $T=2$ case, the Taylor theory fails, but the degree conjecture is upheld, with $T=3$ maintained in the final state. In fact, there are four elliptic periodic orbits and one hyperbolic periodic orbit, so the four initial columns have led to four twisted flux tubes in the final state. The degree constraint itself would not preclude the annihilation of the hyperbolic orbit with one of the elliptic orbits. There must be some further topological constraint on the evolution preventing this. This remains to be fully explored.

In conclusion, our examples show that preservation of the topological degree $T$ leads to non-trivial constraints on braided magnetic flux tubes. We therefore claim that the topological degree imposes a constraint on the relaxation beyond that of helicity conservation. In particular, the Taylor state is not reached if its degree differs from that of the initial state, at least in cases like our examples where the boundary is unaffected by the dynamics. Note that, although these simulations use line-tied boundary conditions (in $z$), the idea extends to the case of a $z$-periodic boundary and hence to a toroidal domain. Finally, we point out that a series of further invariants may be constructed by taking the topological degree of multiple iterations of the field line mapping, that is, by considering the degree of periodic orbits with periods greater than one.


%
%

%

\begin{acknowledgments}
This work was supported by the UK Science \& Technology Facilities Council (grants ST/G002436/1 and PP/E004628/1). Simulations used the UKMHD consortium parallel cluster funded by STFC and SFC (SRIF) at the University of St Andrews.
\end{acknowledgments}

\providecommand{\noopsort}[1]{}\providecommand{\singleletter}[1]{#1}%
%


\end{document}